\documentclass[12pt,a4paper]{article} 
\usepackage{graphicx}
\usepackage{lscape}
\usepackage{amsmath}
\usepackage{subfig}
\usepackage{xcolor}
\usepackage[normalem]{ulem}
\usepackage{slashbox}
\usepackage{multirow}
\usepackage{cite}

\begin{document}

\title{Pantheon update on a model-independent analysis of cosmological supernova data}
\renewcommand{\thefootnote}{\fnsymbol{footnote}}
\author{A. Kaz\i m \c Caml\i bel$^{1}$\footnote{camlibel@tau.edu.tr}, \. Ibrahim Semiz$^{2}$\footnote{ibrahim.semiz@boun.edu.tr} \& Mehmet Akif Feyizo\u{g}lu$^{2}$\footnote{akif.feyizoglu@boun.edu.tr} \\
$^{1}$\small Electrical and Electronics Engineering \\
\small T\"{u}rk-Alman Universitesi (Turkish-German University)\\
\small 34820 Beykoz, \.Istanbul, Turkey  \\
$^{2}$\small Physics Department, Bo\u gazi\c ci University \\ 
\small 34342 Bebek, \.Istanbul, Turkey }
\date{}

\maketitle
\renewcommand{\thefootnote}{\arabic{footnote}}\setcounter{footnote}{0}

\begin{abstract}
We present an update of our previous work, necessitated by availability of a significantly improved dataset. The work is a model-independent analysis of the cosmological supernova (Type Ia) data, where function families are fit to the data in form of luminosity distance as function of redshift, that is, $d_{L}(z)$; and subsequently time-derivatives of the scale function $a(t)$ are {\it analytically} derived, but as functions of $z$, without making assumptions about of gravity or the contents of the universe. This gives, e.g. the redshift value at which the universe goes over from deceleration to acceleration, as $z_{t}=0.54 \pm 0.04$ for a flat universe. In the update, we switch to a more modern fit criterion and also take into account the uncertainty in the calibration of the SNIa luminosities.

If a theory of gravity {\it is} assumed, our results allow determination of the density of the universe as function of $z$, from which conclusions about the contents of the universe can be drawn. We update the previous work's result where this was done for Einstein gravity, finding a lower-limit on the dark energy fraction, $\Omega_{DE}>0.46$;  and here we do this also for Starobinsky gravity, where we can find a Starobinsky parameter that can eliminate the need for dark energy.
\end{abstract}


\section{Introduction}

As is well known, type Ia supernova (SNIa) observations \cite{supernovateam, supernovaproject} are taken as the first and foremost manifestation of accelerated expansion of the universe. In a recent paper \cite{semiz2015cosmological}, we proposed a model-independent method for analysis of cosmological supernova type Ia (SNIa) data, and questioned what they {\it really} tell us about the evolution of cosmic expansion.  The present paper is an update of that work in several respects,  necessitated by the compilation of the Pantheon dataset, containing about twice as many SNIa observations as before, with smaller relative errors in the observations.

Obviously, the analysis of any scientific data requires a number of assumptions that constitute an underlying model. In cosmology, these assumptions  arise in the form of geometry of spacetime (usually isotropy and homogeneity), contents of the cosmic fluid (a subset of radiation, baryonic matter, dark matter, dark energy, etc.) and a theory of gravity (General Relativity or one of its modifications or alternatives). The model seemingly with the minimal set of reasonable assumptions, the model currently assumed correct until proven otherwise, the ``concordance model'', is the so-called $\Lambda$CDM model, which assumes that General Relativity (GR) with cosmological constant $\Lambda$ describes gravitation, and dark matter (DM) exists. The often-heard statement that the universe contains approx. 5 \% normal matter, 25 \% dark matter, and 70\% dark energy (DE, here equivalent to $\Lambda$) is a statement of {\it this} model \cite{aghanim2018planck}.

Yet, one can refrain from making any assumptions on cosmic ingredients or theory of gravity, other than it being a metric theory; and still derive conclusions. These conclusions would be about the spacetime geometry of the universe, hence such an approach is generally termed {\it cosmography} \cite{cosmography} or a {\it model-independent} approach. Our previous paper was in that spirit. 

Our proposal in that work \cite{semiz2015cosmological} was to start with candidate relations that fit luminosity distance ($d_L$) versus redshift ($z$)  of observed SNeIa well (we used the Union 2.1 data set \cite{union21})  and construct the first and second time-derivatives of scale factor using Friedman-Robertson-Walker (FRW) metric, leaving the spatial curvature free within a reasonable range. It should be emphasized that the method is a hybrid numerical/analytical algorithm, since the time-derivatives $\dot{a}$ and $\ddot{a}$ can be found {\it analytically} as a function of $z$ once a numerical fit for the parameters of the $d_L(z)$ candidate function is performed. Another unique aspect of the algorithm is that the number of parameters is also variable; the goodness-of-fit criterion makes the choice.

In \cite{semiz2015cosmological} we found that the result of the analysis with {\it only} the Union 2.1 SNIa data are very sensitive to the choice of the candidate relation, the number of parameters that candidate possesses and even to the different parametrizations of redshift itself. From these data, it is not even obvious that the universe has transitioned from a decelerating phase to an accelerating one. Important parameters such as the current value of acceleration (even its sign) and the transition redshift (deceleration to acceleration, when applicable) are among the results that are not consistent among the different candidate relations. This suggests that previous reconstructions in the literature of cosmic history from analyses of SNIa data within the framework of a model (cosmic fluid properties and gravity theory) are at most as reliable as the model they start from. In our work, we were able to ``tame'' these inconsistencies and attain a consistent picture of cosmic evolution only by including the high redshift gamma ray burst (GRB) data, hence, by slightly compromising the model-independent nature of the study.

For the next stage, we assumed that GR is the correct theory of gravity, but we made no assumptions about the content of the universe (except the observational fact that it contains {\it some} matter). This still allowed calculation of the density of the universe as function of redshift (again analytically), hence a check if matter is enough. It can be seen that dark energy (DE) is inescapable if GR is correct, and an upper-limit can be put on the matter fraction $\Omega_{m}$ in the universe, hence a lower-limit on the DE fraction  $\Omega_{DE}$. 

The present work is an update of the previous one in several respects: (i)~more current, richer and ``better'' dataset (ii) more modern goodness-of-fit criterion (iii) recognition of the uncertainty in the calibration of luminosity distances (iv)~a modified-gravity theory (that of Starobinsky \cite{starobinsky1980new}) in addition to GR in the second stage. The inconclusive nature of the previous analysis of the SNIa observations may be attributed to the highly scattered data compilation with big uncertainties at redshifts $z\sim1$. It is natural to expect that with a higher number of more precise observations and increasing number of higher redshift SNeIa, cosmographic parameters would be better constrained, and SNIa data would start to tell us more. In fact, in the mean time, a more recent compilation, the so-called Pantheon dataset \cite{scolnic2018complete} has become available, making it necessary to rework the analysis with updated data. This compilation consists of 1048 data points up to a redshift $z\sim2.3$ coming from different redshift surveys.

We also decided to use the more modern $\chi^2$-probability criterion to measure goodness-of-fit instead of $\chi^2$-per-degree-of-freedom. Finally, we enriched the second stage by not just using GR to derive an updated upper-limit to the matter content of the universe, but also using Starobinsky gravity to see if dark energy can be made unnecessary in that theory.


\section{The Data and the Method} 

In this section we first compare the dataset of the previous paper, the Union 2.1 dataset, to the current one, the Pantheon dataset. Although the method is described in detail in the previous paper \cite{semiz2015cosmological}, we proceed to give  an outline in the interest of being reasonably self-contained; with the additions necessary for using Starobinsky gravity in the second stage. 

\subsection{Union 2.1 vs. Pantheon data} \label{unionVSpantheon}

\begin{figure}[h!]
\centering
\includegraphics[width=\columnwidth]{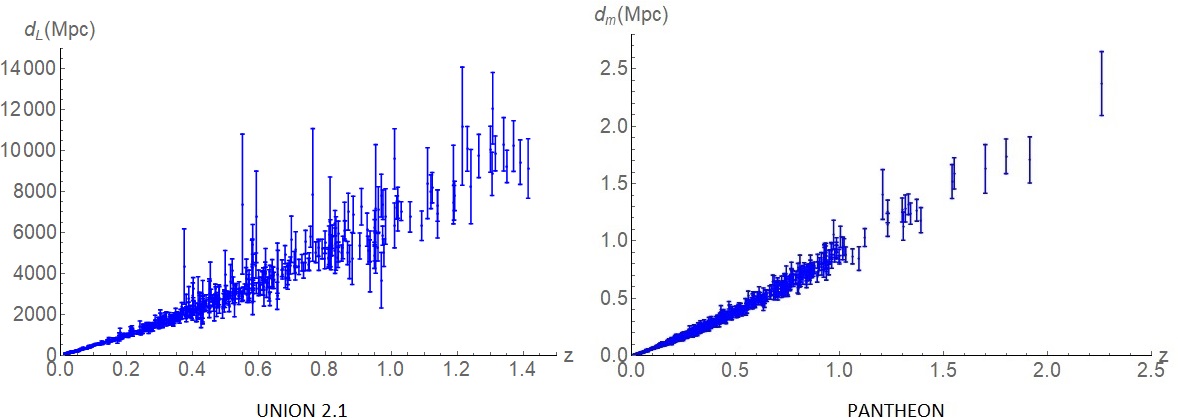}
\caption{Comparison of the Union 2.1 and Pantheon datasets. The different variables on the vertical axes are explained in the text. } 
\label{fig:unionvspantheon} 
\end{figure}

In Fig.\ref{fig:unionvspantheon}, we see the two datasets; note that the vertical axes are different, to be explained below. Of course, the number of data points is larger in the more current dataset, and it extends further, to $z \approx 2.3$ instead of $z \approx 1.5$. But the latter aspect is somewhat misleading, since only six data points out of 1048 are beyond $z \approx 1.5$, and only one beyond $z = 2$. It can also be seen that the error bars have been reduced; this, in combination with larger number of data points  leads to the expectation of more consistent results when the expansion history of the universe is reconstructed.  

The Union 2.1 data are given as ($\mu$, $z$) pairs where $\mu$, the distance modulus, satisfies
\begin{equation}
\mu = m - M,
\label{eq:mu=m-M}
\end{equation}
and $m$ and $M$ are the apparent and absolute magnitudes, respectively. The distance modulus $\mu$ can be converted easily to luminosity distance by 
\begin{equation}
d_L(z) = 10^{\mu/5+1}\;\textnormal{pc} = 10^{\mu/5-5}\;\textnormal{Mpc}  \label{lumdis_mu}
\end{equation}
The data were used in this form in the previous paper \cite{semiz2015cosmological} and this is why the vertical axis of the left panel of Fig.\ref{fig:unionvspantheon} is labeled as $d_L$.   

However, the Pantheon data are given as ($m$, $z$) pairs, since $m$ is what is really measured. These cannot be directly converted to luminosity distance, the value of $M$ is needed (hence the importance of standard candles!). While there is strong empirical evidence and theoretical basis for believing that SNeIa {\it do} constitute standard candles, the calibration, that is, the value of $M$, is not very certain. This calibration also involves $H_{0}$, the current value of the Hubble parameter, and there is a well-known tension \cite{riess20162} between the locally determined value (from Cepheids) and that determined from CMB analysis.

Hence, we decide to factor out the term involving $M$, i.e. define a distance indicator in terms of $m$ only, calling it the ``$M$-transformed luminosity distance'', and denoting it by $d_{m}$: 
\begin{equation}
d_L(z) = 10^{\left(\frac{m-M}{5}-5\right)}\;\textnormal{Mpc} = 10^{-M/5} d_{m} \Longrightarrow d_m \equiv 10^{\left(\frac{m}{5}-5\right)}\: \textnormal{Mpc},
\label{M-lumdis}
\end{equation}
and hence, this variable appears on the right panel of Fig.\ref{fig:unionvspantheon}.

It turns out that the $\mu$ values of the Union 2.1 compilation were obtained by using the fiducial value $M=-19.3$, equivalently $H_{0}$ = 70 $\frac{\textnormal{km/s}}{\textnormal{Mpc}}$, hence our result for $H_{0}$ in \cite{semiz2015cosmological} (after eq.(3.3)), was a tautology/verification, after all.  

\subsection{Redshift, scale factor and luminosity distance} \label{calc}

It is known~\cite{SN1a_std_cdl} that the function $d_L(z)$ is related to the function $a(t)$ by 
\begin{equation}
d_L(z)=(1+z)a_0f_k^{-1}\left(\frac{c}{a_0}\int_{0}^{z}\frac{dz'}{H(z')}\right) \label{eq:lumdisfroma}
\end{equation}
where $H = \dot{a}/a$ is the Hubble parameter and the overdot denotes the time derivative; 
\begin{equation}
f_k(x) = \int_{0}^{x}\frac{dr}{\sqrt{1-kr^2}}=
\begin{cases}
\sin ^{-1}(x) &k=1 \\
x &k=0 \\
\sinh ^{-1}(x) &k=-1, \\
\end{cases}
 \label{f_k}
\end{equation}
$k$ the spatial curvature parameter with $k=1,0$ and $-1$ representing spatially closed, flat and open universes, respectively; and $a_0$ the present value of the scale factor [The contents of the big parenthesis in eq.(\ref{eq:lumdisfroma}) is the argument of the function $f_k^{-1}$, it does not represent a multiplication]. This relation is needed since measurements are $d_L(z)$ points, hence a $d_L(z)$ function is needed to check them against; however a cosmological model usually starts from some assumptions of contents of the universe and law of gravity, and eventually predicts a function $a(t)$.  

The main idea in our approach \cite{semiz2015cosmological} is that the relation (\ref{eq:lumdisfroma}) can be inverted:
\begin{equation}
t_0-t=\int_{0}^{z}\frac{dz}{(1+z)c\sqrt{1-\kappa\frac{d^2_L(z)}{(1+z)^2}}}\frac{d}{dz}\left(\frac{d_L(z)}{1+z}\right)  
\label{lookback}
\end{equation}
where $\kappa$ is defined as
\begin{equation}
\kappa \equiv \frac{k}{a^2_0}
\end{equation}
thus it is a parameter that can take continuous values. We choose 11 equally-spaced values (of course, including 0) between $-\kappa_{0}$ and $+\kappa_{0}$, where $\kappa_{0}=(10000~\rm{Mpc})^{-2}\sim(2c/H_0)^{-2}$, since the SNIa data do not tell anything about curvature.

Eq.(\ref{lookback}) gives $t(z)$ once the function $d_L(z)$ is specified. Inverting that relation would give $z(t)$, and the relation
\begin{equation}
a = \frac{a_{0}}{1+z}, 
\label{a-of-z}
\end{equation}
derivable from the FRW metric and the equivalence principle, would give $a(t)$. However, neither the integration in eq.(\ref{lookback}) nor the inversion $t(z) \longleftrightarrow z(t)$ can be performed analytically except in the simplest cases; hence the argument would seem to be a meaningless exercise in practice. But, consider the time-derivative of $a$:
\begin{equation}
\frac{da}{dt} = \frac{da}{dz} \frac{dz}{dt}.
\end{equation}
Here $da/dz$ can be found from eq.(\ref{a-of-z}) and $dz/dt$ is the algebraic inverse of the integrand of eq.(\ref{lookback}); hence $da/dt$ can ve evaluated analytically in terms of $z$; for any value of $\kappa$, that is, curvature. Similarly, all time derivatives can be evaluated in terms of $z$, in particular, the second and third derivatives that we will need.

On the other hand, in the present work, we use $d_m(z)$  instead of $d_L(z)$. This means that eq.(\ref{lookback}) is replaced by 
\begin{equation}
dt_M=-\frac{dz}{c(z+1)}\frac{1}{\sqrt{1-\kappa10^{-2M/5}\frac{d^2_m(z)}{(z+1)^2}}}\frac{d}{dz}\left(\frac{d_m(z)}{z+1}\right) \label{lookback2}
\end{equation}
after defining $t_M=10^{M/5}t$. Hence we can determine
\begin{equation}
\dot{a}|_M\equiv \frac{da}{dt_M} =\frac{da}{dz}\frac{dz}{dt_M} =10^{-M/5}\dot{a} \label{adot_M}
\end{equation}
and similarly, the higher-order derivatives. Note that to use eq.(\ref{lookback2}) we need a value for $M$; we use $M=-19.3$, a possible ``mistake'' of a few percent will correspond to changing the curvature by a few percent, so it is not crucial. The overall common factor is not very important either; we will be interested in intercepts  (e.g. the redshift at which the universe's expansion switches from deceleration to acceleration) or comparisons of different curves.

\subsection{Density vs redshift: GR and Starobinsky} \label{GRvsStarobinsky}

A {\it metric} theory of gravity relates the spacetime geometry to the contents of spacetime; hence, allows us to draw inferences about the contents of spacetime from any information about its geometry. In particular in cosmology, the contents of the universe is usually assumed to consist of one or more perfect fluids; hence from $\dot{a}$ and higher derivatives we find model-independently as described above, we can find the {\it total} density of the universe as a function of the redshift $z$, if we also assume one of the simpler theories of gravity. Note that no assumption about the equation(s) of state of the fluid(s) is made. 

For example, GR gives
\begin{equation}
H^2+\frac{k c^2}{a^2}=\frac{8\pi G}{3c^2}\rho, \label{friedmann}
\end{equation}
the equation also known as the Friedmann equation, as the 0-0 component of Einstein's equations. This equation was utilized in the previous work \cite{semiz2015cosmological} to show that the evolution of the density of the universe is not compatible with only the matter fluid being present, according to the SNIa data and assuming GR; and furthermore, that the extra content must have dark energy-type behavior. 

In the present work, we redo the GR analysis for the updated data, but we also use one of the simplest alternative gravity theories, that of Starobinsky \cite{starobinsky1980new}. That theory belongs to the class known as $f(R)$ theories, for which the action can be written as
\begin{equation}
S={\int}d^4x\sqrt{-g}f(R), \label{einsteinhilbertS}
\end{equation}
$f(R)=R$ corresponding to GR, $f(R)=-2\Lambda+R$ corresponding to GR with cosmological constant $\Lambda$, and 
\begin{equation}
f(R)=R+{\alpha}R^2
\label{starobinskyL}
\end{equation}
giving the Starobinsky theory of gravity \cite{starobinsky1980new}. Together with a perfect-fluid type energy-momentum tensor
\begin{equation}
T^{\mu\nu}=(\rho+p)u^\mu u^{\nu} + p g^{\mu\nu}
\label{perf_fl_T}
\end{equation}
(and assuming the fluid or fluid mixture is static in FRW coordinates), the equations of motion for the action (\ref{einsteinhilbertS}) give 
\begin{equation}
H^2+\frac{kc^2}{a^2}+\frac{1}{f'}\left[f''H\dot{R}-\frac{c^2}{6}\left(f'R-f\right)\right]=\;\frac{8{\pi}G}{3c^2f'}\rho
\end{equation}
which turns into
\begin{align} \label{asde}
\frac{8{\pi}G}{3c^2}\rho=\left(\frac{\dot{a}}{a}\right)^2&+\frac{6\alpha}{c^2}\left(2\left(\frac{\dot{a}}{a}\right)^2\left(\frac{\ddot{a}}{a}\right)-3\left(\frac{\dot{a}}{a}\right)^4-\left(\frac{\ddot{a}}{a}\right)^2+2\left(\frac{\dot{a}}{a}\right)\left(\frac{\dddot{a}}{a}\right)\right)
\\ \nonumber
&+\frac{kc^2}{a^2}-12\alpha\frac{k}{a^2}\left(\frac{\dot{a}}{a}\right)^2+6\alpha\frac{k^2c^2}{a^4} 
\end{align}
for the Starobinsky $f(R).$ This relation can also be used to find $\rho(z)$, once a good $d_L(z)$ or $d_m(z)$ function is identified. 


\section{Model-independent determination of the expansion history of the universe}

In this section, we report on the update necessitated by the availability of the significantly improved dataset Pantheon, of the model-independent part of the analysis in the previous work \cite{semiz2015cosmological}. Due to the improvement, the SNIa data now suggest much more strongly that the expansion of the universe is accelerating now, and was decelerating in the past.

\subsection{Fitting the M-transformed luminosity distance function}

First, a note about the independent variable is in order: As also discussed in the previous work \cite{semiz2015cosmological}, the traditional redshift variable $z$ is not the only redshift variable available for use; depending on the analysis, it may not even be a good variable to work with because of potential convergence problems beyond $z=1$ of Taylor series of various cosmological functions~\cite{cosmography}, all necessarily related to $a(z)$; see eq.(\ref{a-of-z}). Hence, other redshift variables are also used in the literature, such as 
\begin{eqnarray}
y_1 = \frac{z}{1+z} \qquad y_2 = \arctan{\frac{z}{1+z}} \qquad  y_4 = \arctan{z} \\
y_5 = \ln{(1+z)} \qquad y_6 = u=1+z 
\end{eqnarray}
where $y_1$-$y_4$ were introduced in \cite{cosmography, yvariables}, $y_5$ was introduced in the previous work \cite{semiz2015cosmological} by us (we later noticed that \cite{sutherland2014luminosity} also uses it), and $y_6$ was used in the previous work, since it can surprisingly give different result despite its apparent triviality. We also define $y_{0}=z$.

As candidate functions for $d_m(y_i)$, we use the same 8 families of functions (See Table \ref{introducingfamilies}) as in the previous work \cite{semiz2015cosmological}.  We want to choose simple functions in order to hope to be able to perform the analytical calculations described in subsection \ref{calc}, hence polynomials [with no constant term except when using $y_{6}=u=z+1$, since it is known that $d_{m}(z=0)=0$]. The multiplications by $u=z+1$ in F2, F4, F6 and F8 are motivated by the $z+1$ expressions dividing $d_{L}(z)$ or $d_{m}(z)$ in eqs. (\ref{lookback}) and (\ref{lookback2}); the Pad\'{e} approximants
\begin{equation}
\tilde{P}(y,M,N) = \frac{P_M(y)}{P_N(y)+1}
\end{equation}
by their nicer reported behavior \cite{pade1,pade2,pade3} than polynomials in some fitting situations.
%

\begin{table}
\caption{The 8 different families used in fits. $y$ can be any one of the redshift parameters $y_0$ to $y_6$ (except $y_{3}$), $P_N(y)$ is the $N^{\rm th}$ order polynomial with zero constant term (except when using $y_{6}=u$), $u(y)$ is $(1+z)$ expressed in terms of $y$, $\tilde{P}$($y$, $M$, $N$) is the Pad\'{e} approximant in variable $y$ and orders $M$ \& $N$; and $c$ \& $d$ are constants.}
\centering
\begin{tabular}{| c | c |}
\multicolumn{1}{r}{}\\
\hline
Designation & Function family  \\
\hline
F1 & $P_N(y)$  \\
\hline
F2 & $P_N(y) u(y)$  \\
\hline
F3 & $P_N(y) \exp(c y)$   \\
\hline
F4 & $P_N(y) u(y) \exp(c y) $   \\
\hline
F5 & $P_N(y) \exp(c y + d y^2)$   \\
\hline
F6 & $P_N(y) u(y) \exp(c y + d y^2) $   \\
\hline
F7 &$\tilde{P}$($y$, $M$, $N$)   \\
\hline
F8 & $u(y)$ $\tilde{P}$($y$, $M$, $N$)   \\
\hline
\end{tabular}
\label{introducingfamilies}
\end{table} 

\begin{figure}
\centering
\includegraphics[width=\columnwidth]{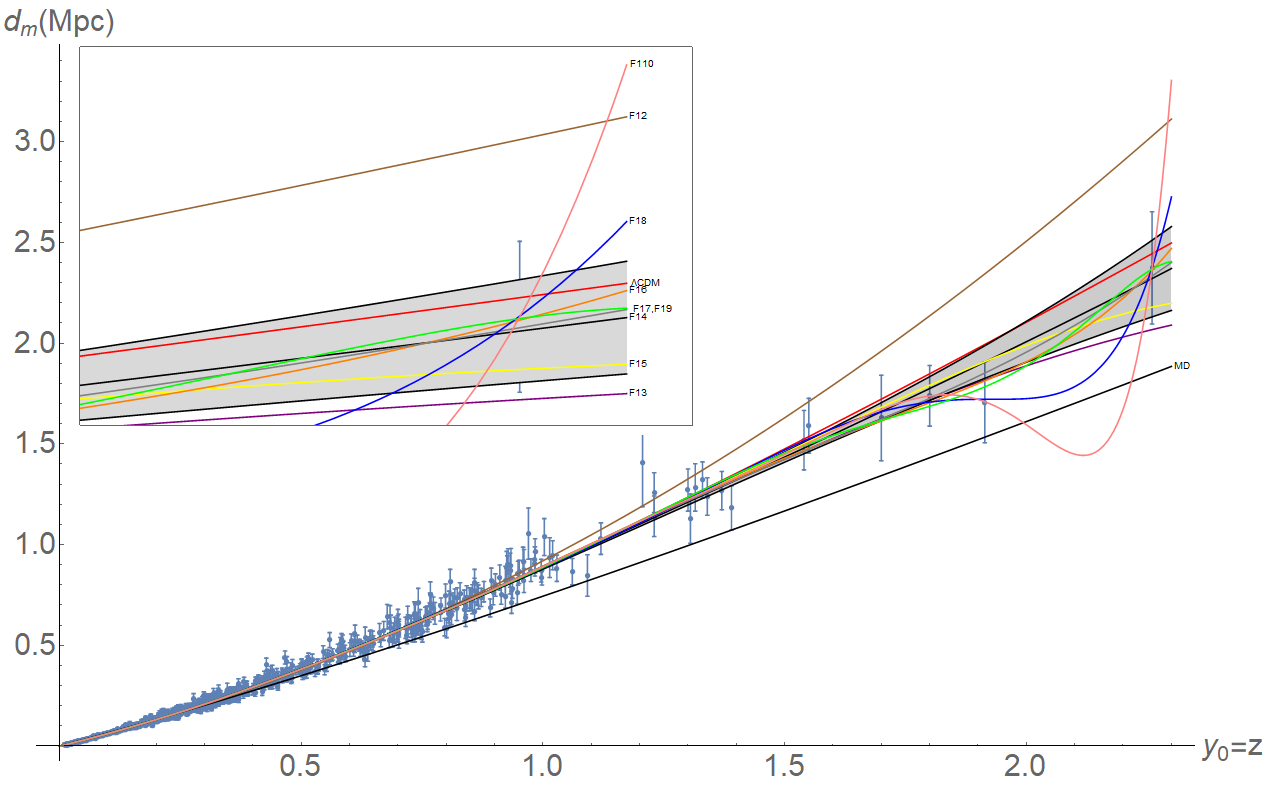}
\caption{Pantheon data (in terms of M-transformed luminosity distance and standard redshift $z \equiv y_{0}$); the N=2 to 10 fits for the first family, i.e. simple polynomials; the fits for the MD (black) and $\Lambda$CDM (red) models, and the one-sigma confidence-levels of the best-fitting member (N=4) of the family F1. The inset shows the right end, magnified. cf. Figure 2 of \cite{semiz2015cosmological}. }
\label{fig:polymodels-y0} 
\thispagestyle{empty}
\end{figure}

In each family one function is chosen as the best fit by looking at the $\chi^2$ probability distribution of the fits (as opposed to $\chi^2$/d.o.f. as in previous work). So, we do not specify the number of free parameters in the beginning, we let the data choose the optimum value. As a demonstration, in Figure \ref{fig:polymodels-y0} we plot the data in terms of the M-transformed luminosity distance and standard redshift $z=y_0$ together with the $N=2$ to $10$ fits for the first family, the one-sigma confidence levels of the best fitting (N=4) member of the first family, the curves of the flat matter-dominated model and the $\Lambda$CDM model. The matter-dominated curve is obviously excluded, and the strange behavior of the high parameter curves ($N=7$ to $10$) can be explained by the overfitting to data. 

\begin{figure}
\centering
\includegraphics[width=\columnwidth]{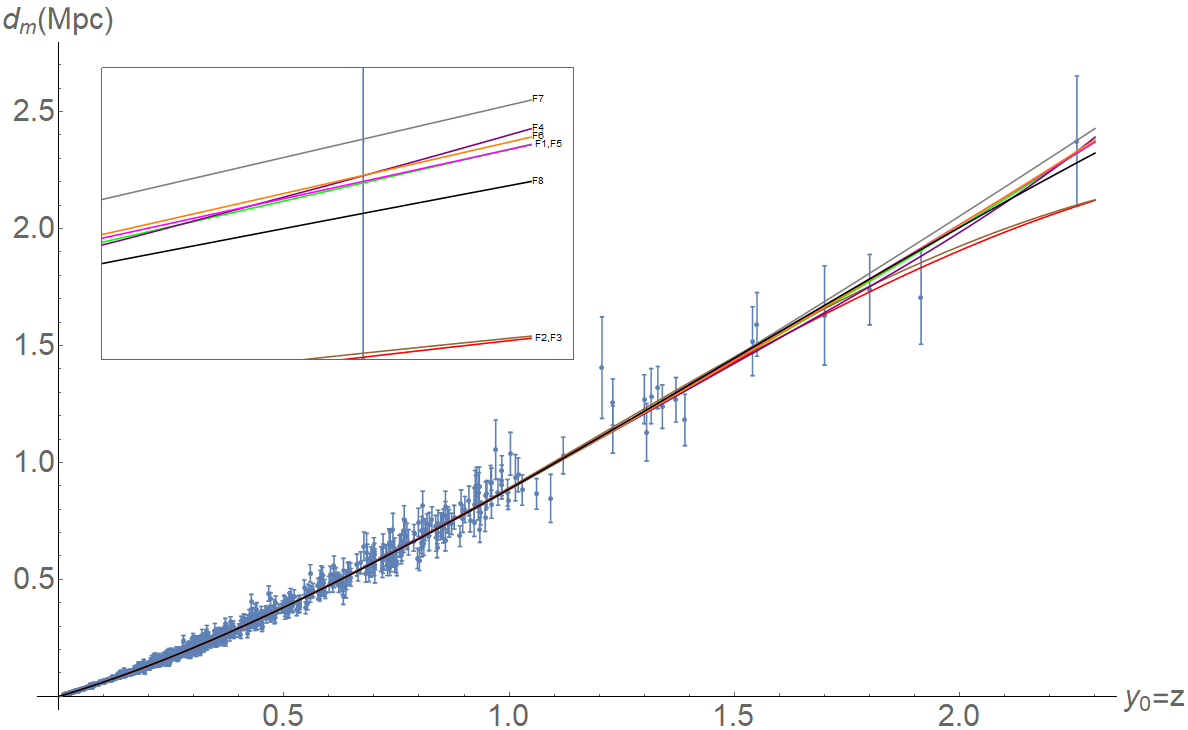}
\caption{Pantheon data (in terms of M-transformed luminosity distance and redshift $z \equiv y_{0}$); and best fits listed in Table \ref{families} for each function family. The one-sigma confidence-levels are not shown to not clutter up the figure; they are similar to those in Figure \ref{fig:polymodels-y0}. cf. Figure 3 of \cite{semiz2015cosmological}.}
\label{fig:bestfits-y0} 
\end{figure}

We show the best fit from each family in Figure \ref{fig:bestfits-y0}, for  $y_0 =z$. This time, we did not include the one-sigma confidence levels of the functions since doing so would render the figure illegible. Similar figures for the other redshift variables are available (SI); no correlation is apparent between the choice of the redshift variable and the pattern of the families' best fits.

All of the best fits of each family and redshift variable, together with their probability values are shown in Table \ref{families}. It seems that only one of the fits (F7,$y_4$) gives a probability value worse than 0.6780, that of $\Lambda$CDM/$y_0$. Hence we will use all of the fits for the time being. It is interesting to note that there are some good fits with as few as two parameters, and also that $y_{0}$ and $y_{6}$ do not always give the same result. 

\clearpage
\begin{table}
\caption{The best fits for the Pantheon SNe Ia data. Each cell displays the internal label(s) of the best-fitting member of the row's family for the column's redshift variable, and the fit's $\chi^2$ probability value. ( The value for $\Lambda$CDM is $0.6780$, using the redshift variable $y_0$, bigger is better).}
\centering
\resizebox{14cm}{!}{
\begin{tabular}{| c | c | c | c | c | c | c |}
\multicolumn{1}{r}{}\\
\hline
\backslashbox{family}{variable}& $y_{0}=z$ & $y_{1}$ & $y_{2}$ & $y_{4}$ & $y_{5}$ & $y_{6}=u$  \\
\hline
F1 & 4; 0.6903 &  4; 0.7310 & 9; 0.7540 & 4; 0.7140 & 3; 0.7012 & 4; 0.6903 \\
\hline
F2 & 2; 0.6854 & 3; 0.6997& 5; 0.7277 & 3; 0.6966 & 3; 0.7027 & 2; 0.6854 \\
\hline
F3 & 5; 0.7044 &  5; 0.7238 & 5; 0.7199 & 5; 0.7075 & 3; 0.7028 & 3; 0.6970 \\
\hline
F4 & 4; 0.6808 &  2; 0.7093 & 5; 0.7280 & 5; 0.7180 & 4; 0.6956 & 4; 0.6808 \\
\hline
F5 & 4; 0.6931 & 5;  0.7220 & 4; 0.7323 & 4; 0.7152 & 3; 0.7002 & 4; 0.6931 \\
\hline
F6 & 4; 0.6890 & 3; 0.7018 & 5; 0.7252 & 6; 0.7084 & 3; 0.7002 & 4; 0.6890 \\
\hline
F7 & 2;1; 0.7024 & 2;1; 0.6936 & 1;2; 0.6982 & 2;1; 0.6389 & 2;1; 0.7005 & 2;1; 0.7024 \\
\hline
F8 & 3;2; 0.7107 & 2;1; 0.7024 & 1;1; 0.7082 & 1;2; 0.6946 & 2;1; 0.6982 & 2;2; 0.6948 \\
\hline
\end{tabular}
}
\label{families}
\end{table}

\subsection{Time derivatives of the scale function}

\begin{figure}
\centering
\includegraphics[width= \columnwidth]{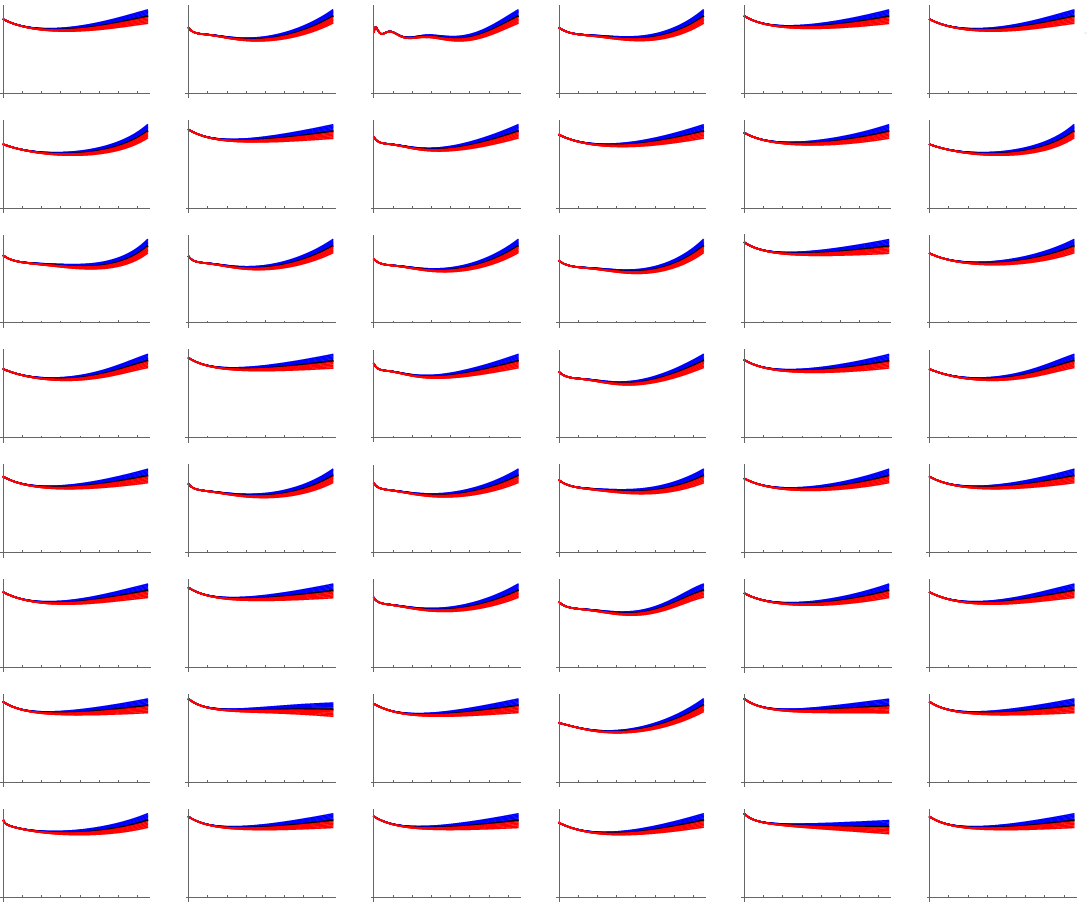}
\caption{The $\dot{a}|_M(z)$ functions, computed analytically for the Pantheon data by eq.(\ref{adot_M}) and similarly for $y_{i}$. The columns represents $y_0$ to $y_6$ (except $y_3$) and rows are for families F1 to F8. Blue, black and red curves are for open, flat and closed spaces respectively. To be able to compare them, horizontal axes are converted to $y_0$, ticked at intervals $\Delta z=0.2$, and the vertical axes are in arbitrary units; cf. Figure $4$ of \cite{semiz2015cosmological}.}  
\label{fig:adotgraph} 
\end{figure}

\begin{figure}
\centering
\includegraphics[width= \columnwidth]{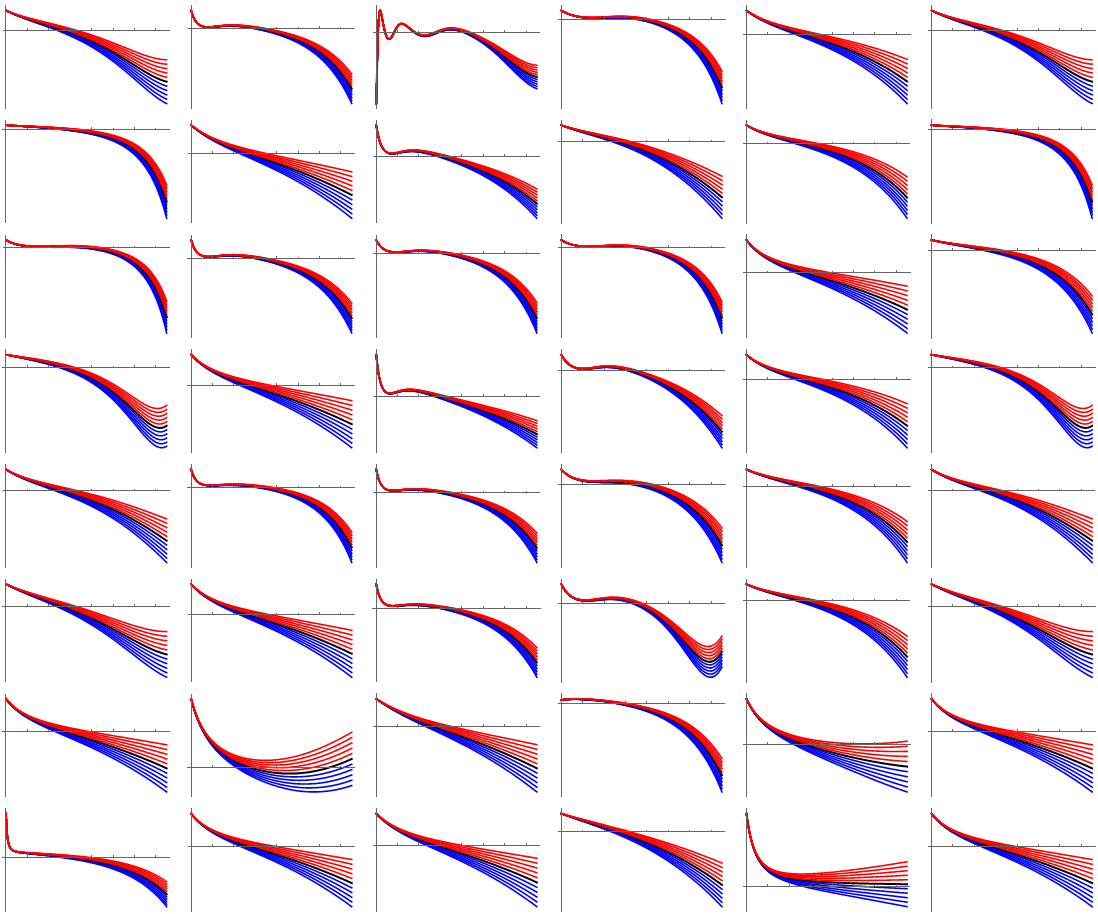}
\caption{The $\ddot{a}|_M(z)$ functions, computed analytically for the Pantheon data by taking $t$-derivative of eq.(\ref{adot_M}) and treating it similarly to express the result in terms of $z=y_0$; and similarly for other redshift variables $y_{i}$. The columns, rows and colors are the same with the previous figure. Again, all horizontal axes are converted to $y_0$, ticked with intervals $\Delta z=0.2$, and the vertical axes are in arbitrary units, hence, these could also be seen as plots of $\ddot{a}(z)$, to be compared to Figure $5$ of \cite{semiz2015cosmological}.}
\label{fig:grid-a-ddot-supernova} 
\end{figure}

\begin{figure}
\centering
\includegraphics[width= \columnwidth]{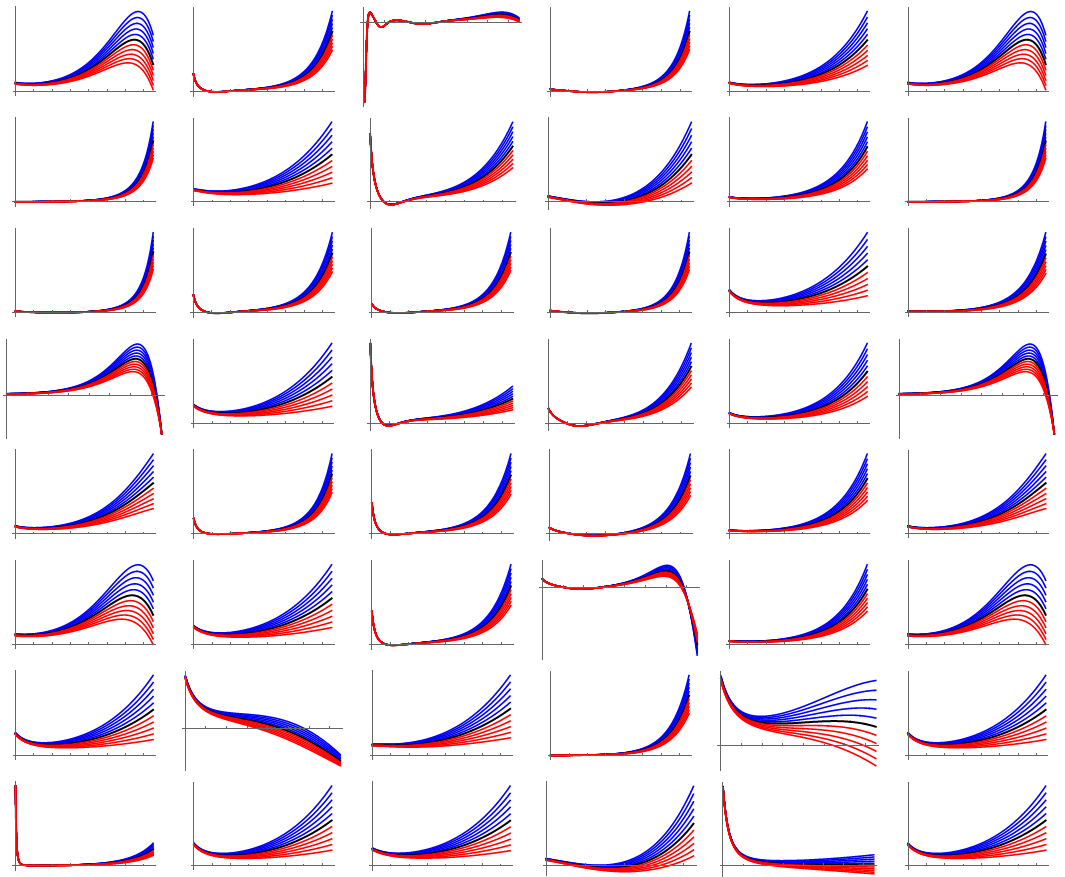}
\caption{The $\dddot{a}|_M(z)$ functions, computed analytically for the Pantheon data by taking two $t$-derivatives of eq.(\ref{adot_M}) and treating it similarly to express the result in terms of $z=y_0$; and similarly for other redshift variables $y_{i}$. The columns, rows and colors are the same as in figures \ref{fig:adotgraph}-\ref{fig:grid-a-ddot-supernova}. Again, all horizontal axes are converted to $y_0$, ticked with intervals $\Delta z=0.2$, and the vertical axes are in arbitrary units, hence, these could also be seen as plots of $\dddot{a}(z)$.}
\label{fig:grid-a-dddot-supernova} 
\end{figure}

The 48 best-fit $d_{m}(z)$ functions (6 variables, 8 families) can now be used to find $a(t)$ numerically, and $\dot{a}(z)$, $\ddot{a}(z)$, etc. analytically, as described in subsection \ref{calc}. As in previous work \cite{semiz2015cosmological}, the 48 numerical plots for $a(t)$ are almost identical, and not very useful in any case, hence we will omit them. We do show the derivatives, in Figs. \ref{fig:adotgraph}-\ref{fig:grid-a-dddot-supernova}. However, we plot them only to $z=1.5$, even though the Pantheon data extend to $z=2.3$. But, as discussed in subsection \ref{unionVSpantheon}, there are only six data points (out of 1048) beyond $z=1.5$; furthermore, the calculations $d_{m}(z) \rightarrow \dot{a}(z), \ddot{a}(z), \dddot{a}(z)$ can introduce singularities around $z \sim 2$ for a few of the resulting functions.

The plots for $\dot{a}$, shown in Figure \ref{fig:adotgraph} and color-coded according to spatial curvature, are not very informative. But the plots for $\ddot{a}$, shown in Figure \ref{fig:grid-a-ddot-supernova}, clearly show periods of past  acceleration and current deceleration, unlike the corresponding figure in \cite{semiz2015cosmological}. The plots for $\dddot{a}$, shown in Figure \ref{fig:grid-a-dddot-supernova} (needed for Starobinsky density analysis), are not very visually informative either.

%
%

Although irregularities of the $\ddot{a}|_M(z)$ are not as bad as they are in the (Union 2.1-based) $\ddot{a}(z)$ functions of the previous work, they are still there. Hence, we will use the same cure, the addition of GRB data \cite{GRB}, even though these are not really suitable for model-independent analysis since cosmological model assumptions are used in the determination of their distance moduli. But they  have large errors compared to Pantheon data, which will give them little weight in the fits. From these low weights and the much smaller number of data, it can be concluded that the GRB data will not model-contaminate the analysis much. 

\begin{figure}
\centering
\includegraphics[width= 0.9 \columnwidth]{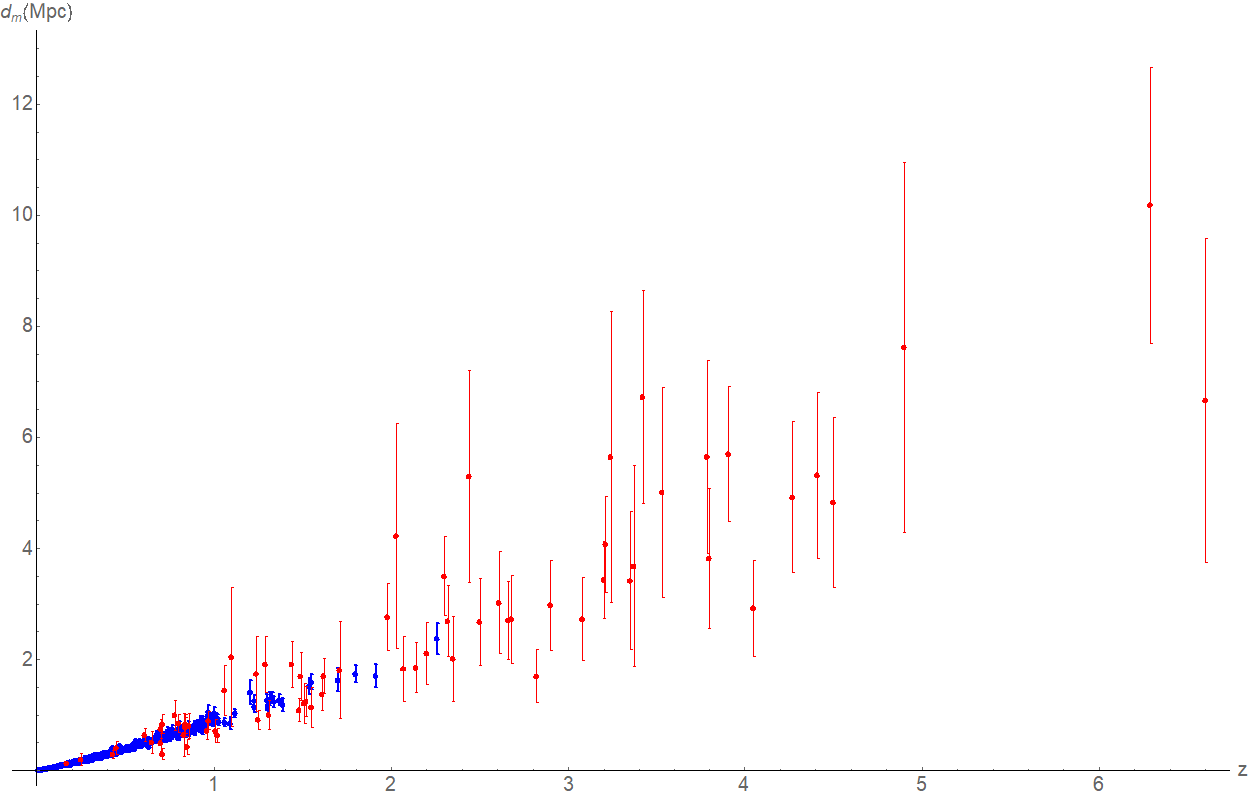}
\caption{The Pantheon supernova data, shown in blue, the GRB data \cite{GRB}, shown in red, together.}
\label{fig:snGRBdata} 
\end{figure}

The GRB data in \cite{GRB} consist of ($\mu$, $z$) pairs, leading to $d_L(z)$. To convert the data to $d_m(z)$ so that they are compatible with our analysis, we use the same procedure as for the Pantheon data, $d_{m}(z) = 10^{M/5} d_L(z)$, as can be seen from eq.(\ref{M-lumdis}). Therefore, we need to use a fiducial value for $M$, we use again $M=-19.3$. But the GRB-data already are model-interpreted, and will only be used to ``tame'' the $\ddot{a}(z)$ functions; so this does not bring appreciable extra model influence. The resulting $d_m(z)$ data for GRBs together with the Pantheon SNIa data are shown in Fig.\ref{fig:snGRBdata}.

\begin{figure}
\centering
\includegraphics[width=\columnwidth]{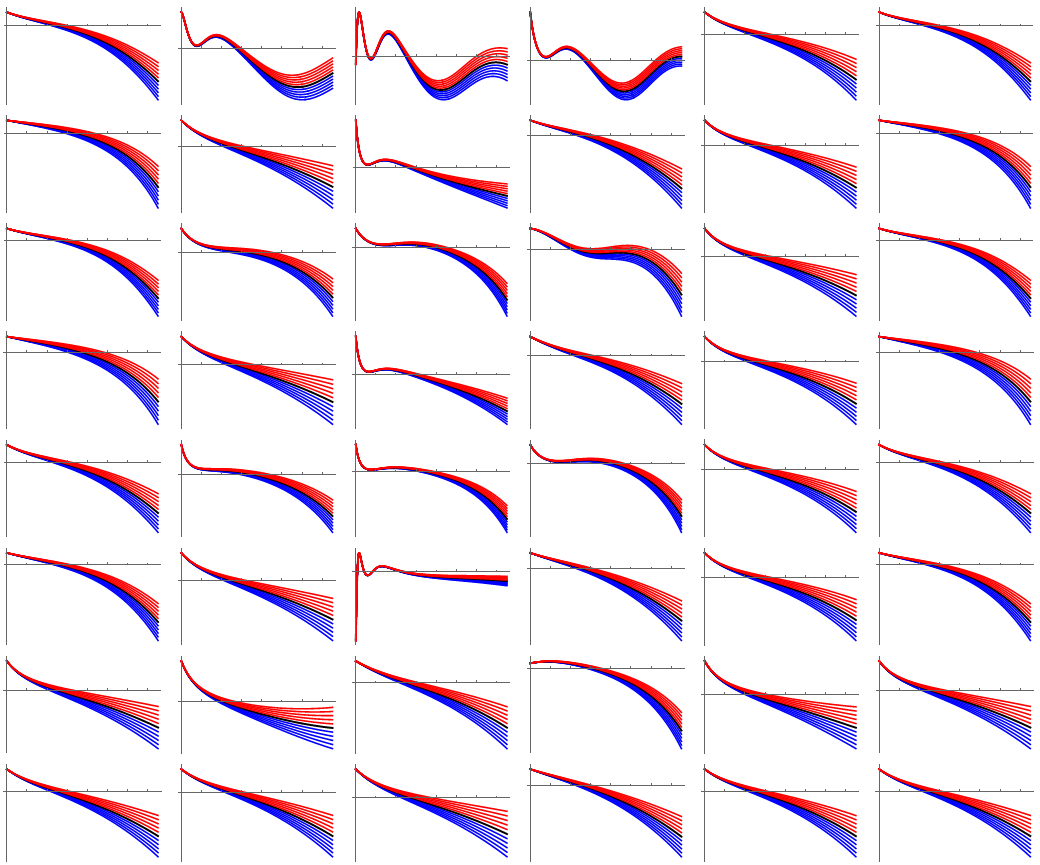}
\caption{The $\ddot{a}|_M(z)$ functions, computed analytically for the Pantheon and GRB data by taking two $t$-derivatives of eq.(\ref{adot_M}) and treating it similarly to express the result in terms of $z=y_0$; and similarly for other redshift variables $y_{i}$. The columns, rows and colors are the same as in figures \ref{fig:adotgraph}-\ref{fig:grid-a-ddot-supernova}. Again, all horizontal axes are converted to $y_0$, ticked with intervals $\Delta z=0.2$, and the vertical axes are in arbitrary units, hence, these could also be seen as plots of $\ddot{a}(z)$, to be compared to Figure 7 of \cite{semiz2015cosmological}.}
\label{fig:grid-a-ddot-supernova-grb} 
\end{figure}

Addition of the 69 GRB data to the 1048 Pantheon supernova data modifies the plots of $\ddot{a}|_M$ to give Figure \ref{fig:grid-a-ddot-supernova-grb}; and the $\chi^2$ probability values of the most useful functions in Table \ref{families2}. Comparing Figure \ref{fig:grid-a-ddot-supernova} to Figure \ref{fig:grid-a-ddot-supernova-grb} indicates that the effect of adding the GRB data is not as dramatic as in previous work, at least for up to $z=1.5$, the range of the plots, as explained in the beginning of this subsection. Still, the number of plots in which we can see the transition from deceleration to acceleration clearly, has increased. Additionally, there is a visual improvement to the plots of columns representing $y_0$, $y_5$ and $y_6$, they do not have any unnatural fluctuations. Strangely, adding the GRB data make columns representing $y_0$, $y_5$ and $y_6$ better but they make columns representing $y_1$, $y_2$ and $y_4$ worse. The functions from family F1 do not give satisfactory result in either case. Another interesting aspect of the GRB addition is that $y_0=z$ and $y_6=z+1$ now give the same curves.
%
\begin{table}
\caption{The most useful fits for the Pantheon SNe Ia + GRB data. Aside from the data used for fitting, everything is same with Table \ref{families}.}
\centering
\resizebox{14cm}{!}{
\begin{tabular}{| c | c | c | c | c | c | c |}
\multicolumn{1}{r}{}\\
\hline
\backslashbox{family}{variable}& $y_{0} = z$ & $y_{1}$ & $y_{2}$ & $y_{4}$ & $y_{5}$ & $y_{6}=u$  \\
\hline
F1 & 5; 06518 & 7; 0.6756 & 7; 0.6751 & 6; 0.6659 & 3; 0.6596 & 5; 0.6518 \\
\hline
F2 & 3; 0.6072 & 3; 0.6584 & 6; 0.6834 & 3; 0.6559 & 4; 0.6537 & 3; 0.6072 \\
\hline
F3 & 6; 0.6443 & 4; 0.6448 & 4; 0.6234 & 4; 0.6112 & 3; 0.6590 & 6; 0.6443 \\
\hline
F4 & 2; 0.5698 & 2; 0.6617 & 5; 0.6865 & 4; 0.6516 & 4; 0.6523 & 2; 0.5698 \\
\hline
F5 & 4; 0.6587 & 4; 0.6573 & 4; 0.6302 & 4; 0.6144 & 4; 0.6526 & 4; 0.6587 \\
\hline
F6 & 4; 0.6518 & 3; 0.6585 & 7; 0.7153 & 3; 0.6557 & 4; 0.6526 & 4; 0.6518 \\
\hline
F7 & 2;1; 0.6578 & 2;1; 0.6252 & 1;2; 0.6580 & 2;1; 0.5284 & 2;1; 0.6505 & 2;1; 0.6578 \\
\hline
F8 & 2;2; 0.6524 & 1;2; 0.6602 & 1;1; 0.6583 & 1;2; 0.6555 & 2;2; 0.6526 & 2;2; 0.6524 \\
\hline
\end{tabular}
}
\label{families2}
\end{table} 

\begin{figure}
\centering
\includegraphics[width= 0.9 \columnwidth]{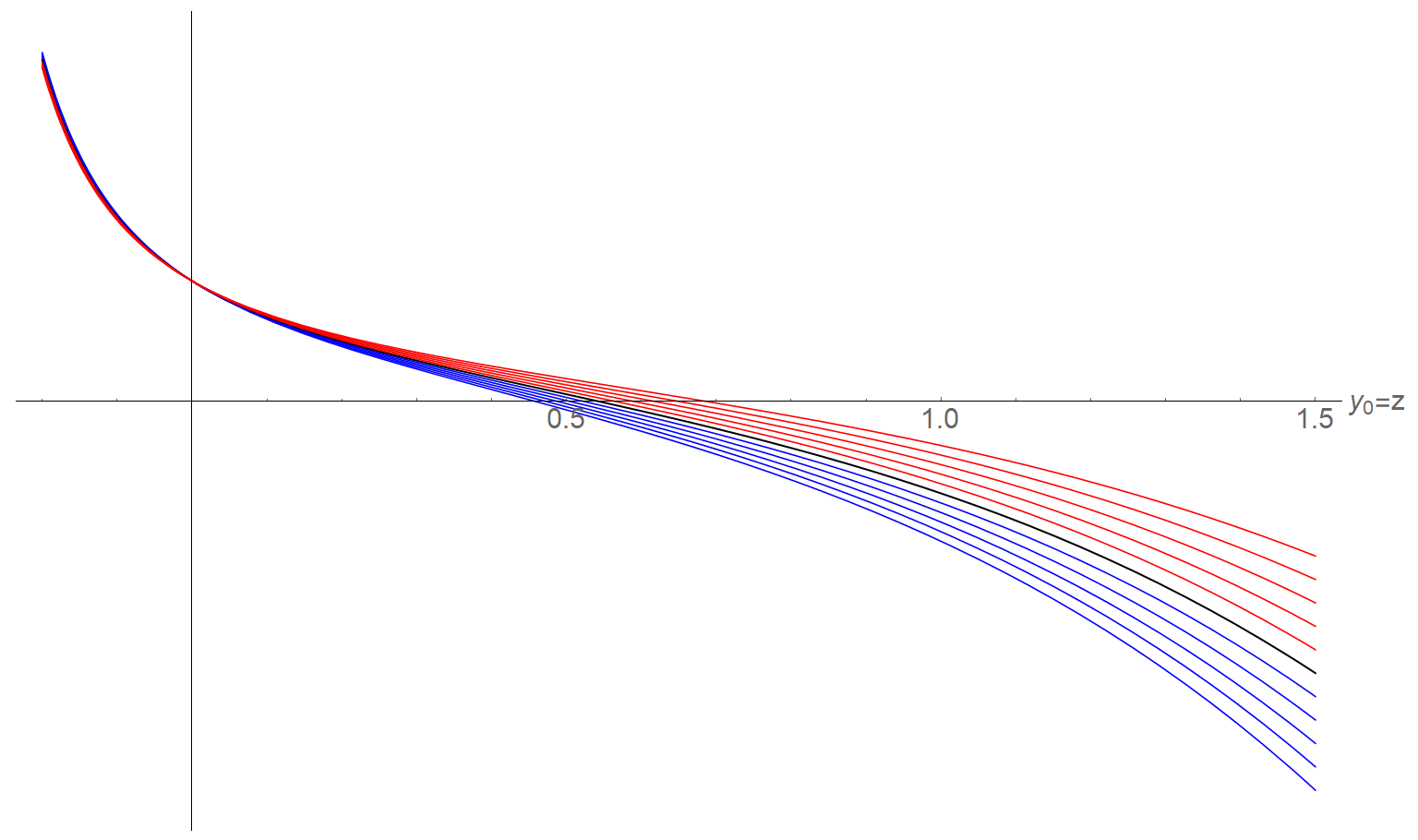}
\caption{The average of the $14$ ``natural'' $\ddot{a}|_M(z)$ functions from Figure \ref{fig:grid-a-ddot-supernova-grb}, namely those in columns one and five; and rows two, three, four, five, six, seven and eight; extended to negative $z$, i.e. to the future. The color coding has the same meaning as in Figure \ref{fig:grid-a-ddot-supernova}. The horizontal axis shows $z=y_{0}$, ticked with intervals $\Delta z=0.1$, and the vertical axis is in arbitrary units.}
\label{fig:supernova-grb-avg-acc} 
\end{figure}

So, we discard the fits belonging to columns two, three, four and six; and also row one of Figure \ref{fig:grid-a-ddot-supernova-grb}. Note that this leaves us with the original $z$ variable and $y_5=\ln(1+z)$. We show the average of the remaining $14$ graphs in Figure \ref{fig:supernova-grb-avg-acc}. Comparing all $14$ plots to this average one by one, we choose the one that is closest to it by minimizing the integral of the absolute value of the difference. The function we find is the one with redshift variable $y5$ and function family F8; we work with this in the next section.

From Figure \ref{fig:supernova-grb-avg-acc}, we can see the redshift of transition from deceleration in the past to current acceleration; for each curvature value. Together with the confidence intervals, the values are
\begin{equation}
z_{t,\textnormal{flat}}=0.54^{+0.04}_{-0.03}   \label{z_tflat}
\end{equation}
for flat universe, 
\begin{equation}
z_{t,+}=0.68^{+0.09}_{-0.06}\quad \textnormal{and} \quad z_{t,-}=0.46^{+0.02}_{-0.02}      \label{z_tcurved}
\end{equation} 
for the most positively and negatively curved universes, respectively.  The corresponding results in the previous work \cite{semiz2015cosmological} were $0.50^{+0.08}_{-0.10}$ for the flat case; and $0.61^{+0.12}_{-0.13}$ and $0.42^{+0.08}_{-0.08}$ for the most positively and negatively curved universes. We can see that the results are consistent, and the confidence intervals similar or somewhat narrower, as expected from the more precise nature of the Pantheon data compared to the Union data. The values (\ref{z_tflat}) and (\ref{z_tcurved}) constitute the most important results of this section.

Results for $z_{t}$ of some previous analyses in the literature are shown in Table \ref{tab:z-trans}. The wide scatter in the results, especially in those that assume some parametrization of a cosmological variable, underscores our point about the reliability of such results, and the importance of model-independent analyses.

\begin{table}[] 
\caption{Results for $z_{t}$ of some previous analyses in the literature. The FLRW metric assumption is common to all analyses, therefore is not stated in the third column; and the date refers to the {\it arXiv} post date. The wide scatter in the results, especially in those that assume some parametrization of a cosmological variable, underscores our point about the reliability of such results, and the importance of model-independent analyses. }
\begin{center}
\begin{tabular}{|c|c|l|c|l|}
\hline
{\bf Date} & {\bf Reference} & {\bf Assumption} & {\bf Dataset} & {\bf Result(s)} \\ \hline
2009 & \cite{guimaraes2009bayesian}   &\multicolumn{1}{l|}{ \begin{tabular}[c]{@{}l@{}}$k=0$ for all\\1) $\Lambda$CDM \\ 2) $q=q_{0}+q_{1}z$ \\ 3) $q$: step at $z_{t}$ \\ 4) jerk $j$ constant \\ 5) $d_L(z)=c_{1}z+c_{2}z^{2}+c_{3}z^{3}$         \end{tabular}} & Union SNIa & \begin{tabular}[c]{@{}l@{}}1) $0.71^{+0.08}_{-0.07}$\\ 2) $0.49^{+0.27}_{-0.09}$\\ 3) $0.46^{+0.40}_{-0.28}$\\ 4) $0.48^{+0.36}_{-0.11}$\\ 5) $0.52^{+0.21}_{-0.08}$\end{tabular} \\ \hline
2015 & \cite{capozziello2015transition} & $f(T)$ gravity, $k=0$ & Union 2.1 & $0.643^{+0.034}_{-0.030}$ \\ \hline
2015 & \cite{rani2015transition} & \multicolumn{1}{l|}{\begin{tabular}[l]{@{}l@{}}$k=0$\\1) $q=q_{1}+q_{2}z$ \\ 2) $q=q_{3}+q_{4}\ln(1+z)$\\ 3) $q=1/2+q_{5}/(1+z)^{2}$  \\ \end{tabular}} & Many & \begin{tabular}[c]{@{}l@{}}1) $0.98$\\ 2) $0.96$\\ 3) $0.60$\end{tabular} \\ \hline
2015 & \cite{semiz2015cosmological} &  \begin{tabular}[c]{@{}l@{}}1) $k=-1$, $a_{0} =$ 10000 Mpc \\ 2) $k=0$\\ 3) $k=1$, $a_{0} =$ 10000 Mpc \end{tabular}& Union 2.1 + GRB &  \begin{tabular}[c]{@{}l@{}}1) $0.42^{+0.08}_{-0.08}$\\ 2) $0.50^{+0.08}_{-0.10}$\\ 3) $0.61^{+0.12}_{-0.13}$\end{tabular}\\ \hline
2016 & \cite{moresco20166l} & \begin{tabular}[l]{@{}l@{}}1) Open $\Lambda$CDM, $H_0=73$\\ 2) None\end{tabular} & $H(z)$ &\begin{tabular}[c]{@{}l@{}}1) $0.64^{+0.11}_{-0.07}$ \\ 2) $0.4^{+0.1}_{-0.1}$\end{tabular}\\ \hline
2016 & \cite{farooq2017hubble} &\begin{tabular}[l]{@{}c@{}}1) $H_0=73$\\ 2) $H_0=68$\end{tabular} &$H(z)$ &\begin{tabular}[c]{@{}l@{}}1) $0.84^{+0.03}_{-0.03}$ \\ 2) $0.72^{+0.05}_{-0.05}$\end{tabular}  \\ \hline
2018 & \cite{refld0} & \multicolumn{1}{l|}{\begin{tabular}[c]{@{}l@{}}$q=a+b(1+z)e^{-z(z+c)}$ \\ $k=0$\end{tabular}} & Pantheon$+H(z)$ & $0.65^{+0.19}_{-0.17}$ \\ \hline
2018 & \cite{aghanim2018planck} & Flat $\Lambda$CDM & Planck & $\approx0.6$ \\ \hline
2019 & \cite{Dinda:2019mev} & \begin{tabular}[l]{@{}l@{}}Complicated\\ parametrization\\ $k=0$\end{tabular} & Many & $\approx0.5$ \\ \hline
2019 & \cite{lin2019non} &$ k=0$ & \begin{tabular}[c]{@{}l@{}}1) Pantheon  \\ 2) Pantheon+$H(z)$\end{tabular} & \begin{tabular}[c]{@{}l@{}}1) $0.59^{+0.08}_{-0.06}$\\ 2) $0.59^{+0.05}_{-0.05}$\end{tabular} \\ \hline
2019 & \cite{goswami2019friedmann} & $a(t)=t^{a}\exp(bt)$ & Planck & $0.956$ \\ \hline
2019 & \cite{jesus2019gaussian} &-& $H(z)$ & \begin{tabular}[c]{@{}l@{}}1) $0.59^{+0.12}_{-0.11}$\\ 2) $0.57^{+0.14}_{-0.10}$\\ 3) $0.58^{+0.13}_{-0.11}$\end{tabular} \\ \hline
2019 & \cite{jesus2019gaussian} &$ k=0$ & Pantheon & \begin{tabular}[c]{@{}l@{}}1) $0.683^{+0.11}_{-0.082}$\\ 2) $0.83^{+0.25}_{-0.50}$\\ 3) $0.69^{+0.23}_{-0.16}$\end{tabular} \\ \hline
2020 & \begin{tabular}[c]{@{}c@{}}present\\ work\\ \end{tabular} &  \begin{tabular}[c]{@{}l@{}}1) $k=-1$, $a_{0} =$ 10000 Mpc\\ 2) $k=0$\\ 3) $k=1$, $a_{0} =$ 10000 Mpc\end{tabular}& Pantheon + GRB &  \begin{tabular}[c]{@{}l@{}}1) $0.46^{+0.02}_{-0.02}$\\ 2) $0.54^{+0.04}_{-0.03}$\\ 3) $0.68^{+0.09}_{-0.06}$\end{tabular}\\ \hline
\end{tabular}\\
\end{center}
* The authors promise to keep an expanded and continuously updated version of this table online for as long as possible and reasonable.
\label{tab:z-trans}
\end{table}

In Figure \ref{fig:supernova-grb-avg-acc}, we also show the part where redshift is negative, predicting that the rate of acceleration will continue to increase in the future. But we do not feel justified to extrapolate very far into the future, given that fits usually behave strangely outside the data interval (and obviously we have no data of the future), so we plot only up to $z=-0.2$. All 14 curves predict an increase in the rate of acceleration; some moderately, some more sharply. We can alternatively display the acceleration in terms of the so-called deceleration parameter $q=-\ddot{a}a/\dot{a}^{2}$, 
(SI) which shows a somewhat softer trend; changing to about -1.5 from a current value of about -0.75.

\section{Inferences from SNIa data together with a gravity theory}

As discussed in subsection \ref{GRvsStarobinsky}, in FRW/perfect fluid(s) cosmology with a metric theory of gravity, the density of the universe can be expressed in terms of the scale factor $a(t)$ and its time-derivatives. But once a good $d_m(z)$ function is known, these are known as function of $z$ or alternatives, after a simple conversion using eq.(\ref{adot_M}) and analogs. Hence,  $\rho(z)$ can be found, and from its form, some conclusions can be drawn about the contents of the universe. Below, we apply this to GR and Starobinsky gravity; using the equations in subsection \ref{GRvsStarobinsky}.

\subsection{Inference with Einstein gravity (GR)}

\begin{figure}
\centering
\includegraphics[width= 0.95 \columnwidth]{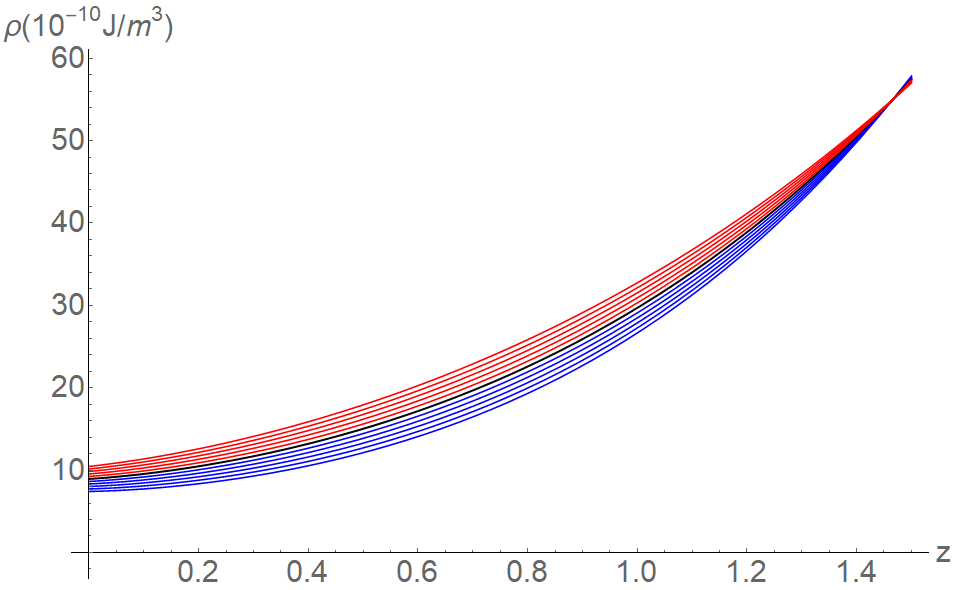}
\caption{The density of the universe as function of $z$, as calculated using the redshift variable $y5$ and fit family F8, assuming Einstein gravity. The color-coding is the same as the one in Figure \ref{fig:adotgraph}. Note the intersection around $z \approx 1.5$.}
\label{fig:rho_y5_f8} 
\end{figure}
Eq.(\ref{friedmann}) gives Figure \ref{fig:rho_y5_f8} for  the redshift variable $y5$ and function family F8, which is closest to the average, as stated in the previous section; again, the confidence intervals are not indicated to not clutter the figure, and the $z$-range has been taken as 0-1.5 for reasons explained before. Even with the different dataset and function used, Figure \ref{fig:rho_y5_f8} is almost same with the corresponding figure of the previous work.

Note that in Figure \ref{fig:rho_y5_f8} there is a specific value of redshift where the density is independent of the spatial curvature, i.e. there is an intersection point of all the plots, as in previous work \cite{semiz2015cosmological}.  The analytical condition for the existence of this intersection point, called $z_*$, was derived in the previous work; it is not a condition that is obvious and it is not necessarily always satisfied. The significance of this point is not clear; in previous work we used this point to derive conclusions aobout the contents of the universe, however it turns out that this is not exactly justified for some curvatures, in particular, its negative values.

\begin{figure}
\centering
\includegraphics[width = \columnwidth]{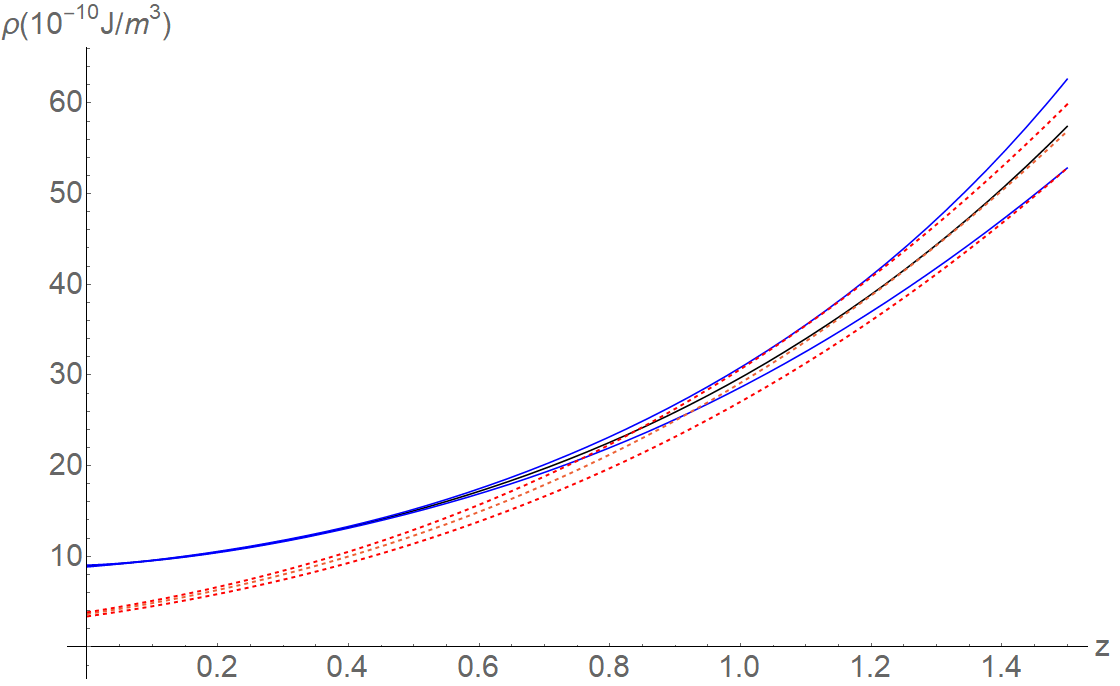}
\caption{The density of the universe as function of $z$, as calculated using the redshift variable $y5$ and fit family F8, assuming GR and spatial flatness (black); together with the confidence interval (blue curves) and the three matter-only curves (dotted red) constrained by the corresponding density curves. The intercepts give the corresponding upper limits to the current matter density, leaving a gap to be explained, presumably by dark energy.}
\label{fig:rho_y5_f8-matter} 
\end{figure}

Nevertheless, conclusions {\it can} be drawn, but without using the intersection point. Assuming all contributions to the energy density of the universe are positive, a matter-only curve, proportional to $(1+z)^3$, that stays barely below the determined density curve will represent an upper limit for the matter content. The intercept of the said curve will give the upper limit for the current value of matter density of our universe. For example, Fig. \ref{fig:rho_y5_f8-matter} shows the situation for the $y5$-F8-flat case. We construct similar graphs and get limits for the most positive and negative values considered of curvature, too; and repeat for the 14 choices of redshift parameter and function family, giving a total of 42 cases. 

The difference between the determined density of the universe and matter density can be ascribed to dark energy, since it starts to dominate as the present is approached. Therefore the difference of the intercepts in Fig.\ref{fig:rho_y5_f8-matter} (or any one of the 42 analogous figures) represents a lower-limit to the dark energy density today. Dividing each density limit by the critical density, that is, the determined density for the corresponding flat case, we can express these upper/lower limits as $\Omega$'s. The 42 calculated upper/lower limits are shown in Table \ref{table:OmegaLims}. 
\begin{table}[]
\caption{$\Omega_{0m,{\rm max}}$'s (upper values in cells) and $\Omega_{0DE,{\rm min}}$'s (lower values in cells) with their estimated errors calculated for 42 different $d_m(z)$ functions (two variables, three curvatures and seven families); assuming GR. Note that the flat $\Omega$'s add up to 1, the others don't.} 
\centering
\begin{tabular}{|l|l|l|l|l|l|l|}
\hline
\multirow{2}{*}{}& \multicolumn{3}{c|}{$y_0$} & \multicolumn{3}{c|}{$y_5$} \\ \cline{2-7} 
&$ k'=-1$ &$ k'=0$ &$ k'=1$ & $k'=-1$ & $k'=0$ &$ k'=1$ \\ \hline
\multirow{2}{*}{F2}  & $0.338^{+0.012}_{-0.012}$ &$ 0.377^{+0.015}_{-0.015}$ & $0.392^{+0.019}_{-0.019}$& $0.368^{+0.015}_{-0.017}$ & $0.401^{+0.026}_{-0.028}$ & $0.400^{+0.029}_{-0.026}$   \\ \cline{2-7} 
 &$0.488^{+0.018}_{-0.017}$  &$ 0.623^{+0.021}_{-0.021}$&$0.782^{+0.024}_{-0.025}$  &$0.463^{+0.027}_{-0.025}$  &$0.599^{+0.038}_{-0.036}$  &$0.769^{+0.036}_{-0.039}$  \\ \hline
\multirow{2}{*}{F3} &$0.369^{+0.013}_{-0.013}$  & $0.426^{+0.016}_{-0.019}$ &$0.460^{+0.024}_{-0.034}$  &$0.367^{+0.014}_{-0.017}$  &$0.390^{+0.028}_{-0.028}$  &$0.390^{+0.028}_{-0.024}$   \\ \cline{2-7} 
 &$0.460^{+0.024}_{-0.023}$ & $0.574^{+0.029}_{-0.026}$ &$0.711^{+0.044}_{-0.034}$  &$0.464^{+0.027}_{-0.023}$  &$0.610^{+0.038}_{-0.038}$   &$0.779^{+0.033}_{-0.038}$  \\ \hline
\multirow{2}{*}{F4} &$0.327^{+0.012}_{-0.012}$   &  $0.357^{+0.015}_{-0.015}$ & $0.365^{+0.016}_{-0.016}$  &$0.367^{+0.014}_{-0.017}$   &$0.396^{+0.023}_{-0.027}$   & $0.395^{+0.026}_{-0.024}$  \\ \cline{2-7} 
 &$0.499^{+0.018}_{-0.018}$  &$0.643^{+0.020}_{-0.021}$ &$0.809^{+0.022}_{-0.021}$  &$0.464^{+0.027}_{-0.025}$   &$0.604^{+0.038}_{-0.034}$  &$0.775^{+0.034}_{-0.037}$  \\ \hline
\multirow{2}{*}{F5} &$0.369^{+0.013}_{-0.014}$    & $0.418^{+0.019}_{-0.023}$   &$0.433^{+0.033}_{-0.032}$    &$0.367^{+0.014}_{-0.015}$    &$0.397^{+0.023}_{-0.027}$    & $0.397^{+0.028}_{-0.025}$   \\ \cline{2-7} 
 & $0.461^{+0.023}_{-0.022}$ &$0.582^{+0.033}_{-0.029}$  &$0.737^{+0.041}_{-0.042}$  &$0.464^{+0.026}_{-0.025}$  &$0.603^{+0.038}_{-0.034}$  &$0.772^{+0.036}_{-0.039}$  \\ \hline
\multirow{2}{*}{F6} &$0.367^{+0.011}_{-0.012}$   &$0.422^{+0.015}_{-0.020}$   &$0.452^{+0.027}_{-0.031}$   &$0.367^{+0.014}_{-0.015}$   &$0.398^{+0.021}_{-0.028}$   &$0.397^{+0.028}_{-0.025}$  \\ \cline{2-7} 
 & $0.462^{+0.021}_{-0.020}$ & $0.578^{+0.029}_{-0.024}$ &$0.719^{+0.040}_{-0.036}$  &$0.464^{+0.026}_{-0.025}$   & $0.602^{+0.039}_{-0.032}$ &$0.772^{+0.036}_{-0.039}$  \\ \hline
\multirow{2}{*}{F7} & $0.366^{+0.015}_{-0.016}$  & $0.387^{+0.027}_{-0.025}$  & $0.387^{+0.024}_{-0.021}$  & $0.372^{+0.012}_{-0.019}$  & $0.385^{+0.031}_{-0.027}$  & $0.386^{+0.027}_{-0.024}$  \\ \cline{2-7} 
 & $0.465^{+0.026}_{-0.025}$  &$0.613^{+0.035}_{-0.037}$  &$0.782^{+0.031}_{-0.034} $ &$ 0.460^{+0.028}_{-0.021} $ &$0.615^{+0.036}_{-0.040}$  &$0.782^{+0.033}_{-0.036}$  \\ \hline
\multirow{2}{*}{F8}& $0.371^{+0.013}_{-0.014}$  & $0.408^{+0.022}_{-0.032}$ & $0.409^{+0.036}_{-0.032}$  & $0.371^{+0.014}_{-0.015}$  & $0.407^{+0.021}_{-0.029}$  & $0.408^{+0.033}_{-0.029}$   \\ \cline{2-7} 
 & $0.459^{+0.025}_{-0.024}$  & $0.592^{+0.043}_{-0.033}$ &$0.761^{+0.043}_{-0.047}$  & $0.459^{+0.026}_{-0.024}$ &$0.593^{+0.040}_{-0.032}$  &$0.762^{+0.040}_{-0.044}$  \\ \hline
\end{tabular}
\label{table:OmegaLims}
\end{table}

To summarize, the Pantheon data tell us that dark energy, whatever its EoS, {\it must} exist, if GR is correct.

\subsection{Inference with Starobinsky gravity}

We would like to know if we can by adopting Starobinsky gravity understand the evolution of the universe without recourse to dark energy, i.e. using only matter; hence we are interested in how well the  density curve $\rho (\alpha,z)$ determined from eq.(\ref{asde}) can match a matter-only curve, i.e. a curve proportional to $(1+z)^3$. In this subsection we investigate this for the spatially flat case, using the variable $y_5$ and the function family F8.

To find the best-matching matter-only curve, we minimize the area between said curves, i.e.
\begin{equation}
\Delta = \int | \rho (\alpha,z)-\zeta\rho(\alpha,0) \; (1+z)^3 | dz \label{rho-matter_Delta}
\end{equation}
where we have parametrized the coefficient of the matter term by the variable $\zeta$ and in terms of the current energy density of the universe. 

\begin{figure}
\centering
\includegraphics[width= 0.95 \columnwidth]{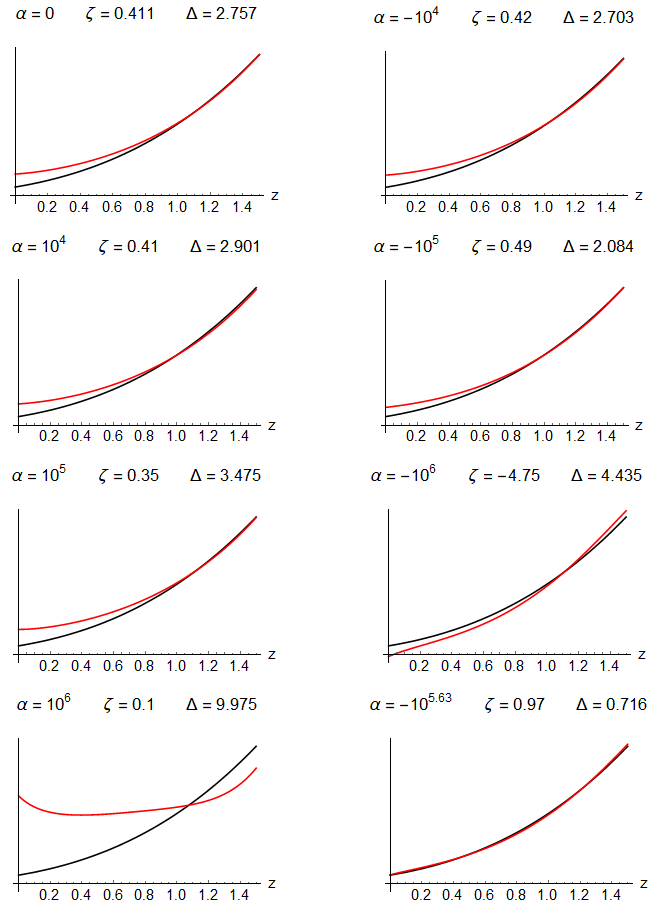}
\caption{$\rho(z)$ (red) and best-fitting matter-only (black) density curves for flat cases with different $\alpha$ values for  Starobinsky gravity. $\zeta$ values are those that minimize $\Delta$, the fitness criterion given by  (\ref{rho-matter_Delta}) for each $\rho(z)$ graph; and the confidence intervals for $\rho(z)$ are not shown. The upper left graph shows the GR case, hence is the same as the black curves in Figs.(\ref{fig:rho_y5_f8}) and (\ref{fig:rho_y5_f8-matter}), the lower right graph shows the case for $\alpha$ that can account for the observed acceleration of the universe.}
\label{fig:rho_matter_many} 
\end{figure}

For $\alpha$ values smaller than $\sim 10^3$ Mpc$^2$, the $\rho (\alpha,z)$ curve is not significantly different from the $\alpha=0$ case ; hence, the optimal $\Delta$ value is not different either, i.e. not good: It was seen in the previous section that no GR/matter-only model can fit our universe. Larger positive $\alpha$ values do not give good results either, and around $\alpha \sim 10^6$ the density starts to show an unacceptable decrease with $z$, so we are led to consider also negative values of the Starobinsky parameter $\alpha$, even though it usually taken to be positive. And for one particular negative $\alpha$ value, $-10^{5.63}$ Mpc$^2$, we {\it do} find a good match. Fig.\ref{fig:rho_matter_many} (again, without confidence intervals to not clutter up the figure), illustrates this result, showing the density curves for different positive $\alpha$ values, together with the matter-only curves (and parameters $\zeta$) that minimize $\Delta$, the integral (\ref{rho-matter_Delta}). The density parameter $\Omega$, defined as the ratio of the density of the universe (i.e. matter) to that of the flat GR universe of same Hubble parameter, is 0.52 for this solution. 

\section{Discussion and Summary}

In this work, we have updated our previous work where we had derived conclusions about the history of expansion of the universe using the SNIa standard candle observations, as much as possible without assuming a theory of gravity or knowledge about the contents of the universe. The reanalysis was needed because a new and improved dataset, the Pantheon one, became available recently.

We also switched to a more modern goodness-of-fit criterion for the fits we needed to perform on the data and to a distance measure independent of the calibration of the SNe Ia. We find analytical expressions for the expansion rate and acceleration of the universe as function of the redshift parameter $z$ or its alternatives, the ones leading to best results being the original $z$ and $y_5=\ln(1+z)$ introduced by us in the previous work. Possibly the most important lesson from our work here is that at the current level of precision of SNIa data, the results for the expansion history of the universe are strongly dependent on the assumed parametrization; results of some parametrizations even contradicting other results in various ways. This casts strong doubt on any models that start from an {\it ad hoc} parametrization of some cosmological variable, without good physical motivation; and then claim good agreement with the SNIa data.

In this work we work with a wide range of parametrizations, in fact, optimizing the parametrizations in subsets, i.e. choosing the best-fitting members of fits of families. We then use  naturalness criteria to eliminate more parametrizations, leading to reasonably consistent results. The important single-number result of this analysis is the transition redshift, i.e. the redshift at which the acceleration of the universe goes over from being negative to positive. We find $z_t = 0.54^{+0.04}_{-0.03}$, consistent with the result in previous work, $0.50^{+0.08}_{-0.10}$. Since the SNIa data do not give us information about the spatial curvature of the universe, we leave that free within a reasonable range, and report similar results for the ends of the considered range. We also suggest that the acceleration will continue to increase for the cosmologically near future, an assertion that also applies in terms of the deceleration parameter.

Of course, an intermediate approach is also possible, assuming a theory of gravity after all, but still refraining from assumptions about the content of the universe, in fact, deriving conclusions about the same. We did this with GR in previous work, showing that in our framework also, a universe containing only matter is incompatible with the data; and deriving lower-limits for the present density for the balance that must exist --dark energy--, without any assumption about its EoS. In this work, we repeat the analysis with  the new data, refining the results; furthermore, we consider another theory of gravity, that of Starobinsky, and ask if the acceleration of the universe can be understood within this modified-gravity formalism without dark energy. The answer we find is negative for positive values of the Starobinsky parameter $\alpha$ (the usual assumption, apparently)  \cite{carvalho2008cosmological}; however, for a particular {\it negative} $\alpha$, we are able to find a good fit to the data, i.e. `explain' the acceleration of the universe. 

The value of this latter excercise can be rightly questioned. After all, it amounts to replacing one parameter, $\Lambda$, of the $\Lambda$CDM model with another parameter, the above-mentioned $\alpha$. The cosmological constant $\Lambda$ has a so-called ``naturalness problem'' if it is interpreted as the quantum zero-point energy of the vacuum, but what if it is interpreted as simply the zeroth term in the Taylor expansion of the function $f(R)$ in eq.(\ref{einsteinhilbertS}) \cite{dadhich2011enigmatic} ?  Would it not then be fair to consider it more natural than $\alpha$? On  the other hand, quintessence (constant EoS parameter $w$) or interacting DE-DM models involve more parameters, hence a Starobinsky model would seem to be simpler than these in some sense. Clearly, which way Occam's Razor will cut is to some extent a matter of taste. 

{\it New} observations in astrophysics and cosmology often bring unexpected discoveries, while richer and more precise observations often allow conjectures to be turned into conclusions or parameters of a paradigm to be sharpened. The Pantheon compilation of the SNIa data belongs to the latter category, and to summarize our work here, we find that the acceleration of the expansion of the universe can be seen model-independently in the Pantheon data significantly clearer than in the Union 2.1 data. It is possible to reasonably reconstruct the expansion history in terms of the redshift, including the redshift value at which the acceleration turned from negative  to positive; and also arrive at conclusions about the contents of the universe or the theory parameter(s) for two of the simplest gravity theories.


\begin{thebibliography}{99}

\bibitem{supernovateam} AG Riess, AV Filippenko, P Challis {\it et al.} ({\it High-z Supernova Search Team}), ``Observational evidence from supernovae for an accelerating universe and a cosmological constant'', {\it The Astronomical Journal} {\bf 116}, 1009 (1998).

\bibitem{supernovaproject} S Perlmutter, G Aldering, G Goldhaber {\it et al.} ({\it Supernova Cosmology Project}), ``Measurements of $\Omega$ and $\Lambda$ from 42 high-redshift supernovae'', {\it The Astrophysical Journal} {\bf 517}, 565 (1999).

\bibitem{semiz2015cosmological} \.{I} Semiz and AK \c{C}aml{\i}bel, ``What do the cosmological supernova data really tell us?'', {\it Journal of Cosmology and Astroparticle Physics} {\bf 2015}, 038 (2015).

\bibitem{aghanim2018planck} N Aghanim, Y Akrami, M Ashddown {\it et al.} ({\it Planck Collaboration}), ``Planck 2018 results. VI. Cosmological parameters'', {\it arXiv}:1807.06209 (2018).

\bibitem{cosmography} C Catto{\"e}n and M Visser, ``Cosmography: Extracting the Hubble series from the supernova data'', {\it arXiv:} gr-qc/0703122 (2007).

\bibitem{union21} N Suzuki, D Rubin, C Lidman {\it et al}, ``The Hubble Space Telescope Cluster Supernova Survey. V. Improving the dark-energy constraints above $z > 1$ and building an early-type-hosted supernova sample'', {\it The Astrophysical Journal} {\bf 746}, 85 (2012).

\bibitem{starobinsky1980new} AA Starobinsky, ``A new type of isotropic cosmological models without singularity'', {\it Physics Letters B} {\bf 91}, 99 (1980).

\bibitem{scolnic2018complete} D Scolnic, D Jones A. Rest {\it et al}, ``The complete light-curve sample of spectroscopically confirmed sne ia from pan-starrs1 and cosmological constraints from the combined pantheon sample'', {\it The Astrophysical Journal} {\bf 859}, 101 (2018).

\bibitem{riess20162} AG. Riess, LM Macri, SL Hoffmann {\it et al}, ``A 2.4 {\%} determination of the local value of the Hubble constant'', {\it The Astrophysical Journal} {\bf 826}, 56 (2016).

\bibitem{SN1a_std_cdl} S Weinberg, {\it Cosmology}. Oxford Univ. Press, 2008.

\bibitem{yvariables} A Aviles, C Gruber, O Luongo and H. Quevedo, ``Cosmography and constraints on the equation of state of the Universe in various parametrizations'', {\it Physical Review D} {\bf 86}, 123516 (2012).

\bibitem{sutherland2014luminosity} W Sutherland and P Rothnie, ``On the luminosity distance and the epoch of acceleration'', {\it Monthly Notices of the Royal Astronomical Society} {\bf 446}, 3863 (2014).

\bibitem{pade1} C Gruber and O Luongo, ``Cosmographic analysis of the equation of state of the universe through Pad{\'e} approximations'', {\it Physical Review D} {\bf 89}, 103506 (2014).

\bibitem{pade2} A Aviles, A Bravetti, S Capozziello and O Luongo, ``Precision cosmology with Pad{\'e} rational approximations: Theoretical predictions versus observational limits'', {\it Physical Review D} {\bf 90}, 043531  (2014).

\bibitem{pade3} H Wei, X-P Yan and Y-N Zhou, ``Cosmological applications of Pad{\'e} approximant'', {\it Journal of Cosmology and Astroparticle Physics} {\bf 2014}, 045 (2014).

\bibitem{GRB} BE Schaefer, ``The Hubble diagram to redshift $>$ 6 from 69 gamma-ray bursts'', {\it The Astrophysical Journal} {\bf 660}, 16 (2007).

\bibitem{guimaraes2009bayesian} A Guimaraes, J Cunha and J Lima, ``Bayesian analysis and constraints on kinematic models from union SNIa'', {\it Journal of Cosmology and Astroparticle Physics} {\bf 2009}, 010 (2009).

\bibitem{capozziello2015transition} S Capozziello, O Luongo and EN Saridakis, ``Transition redshift in $f(T)$ cosmology and observational constraints'', {\it Physical Review D} {\bf 91}, 124037 (2015). 

\bibitem{rani2015transition} N Rani, D Jain, S Mahajan, A Mukherjee and N Pires, ``Transition redshift: new constraints from parametric and nonparametric methods'', {\it Journal of Cosmology and Astroparticle Physics} {\bf 2015}, 045 (2015).

\bibitem{moresco20166l} M Moresco, L Pozzetti, A Cimatti {\it et al.} ``A 6\% measurement of the Hubble parameter at $z\sim 0.45$: direct evidence of the epoch of cosmic re-acceleration'', {\it Journal of Cosmology and Astroparticle Physics} {\bf 2016}, 014 (2016).

\bibitem{farooq2017hubble} O Farooq, FR Madiyar, S Crandall and B Ratra, ``Hubble parameter measurement constraints on the redshift of the deceleration–acceleration transition, dynamical dark energy, and space curvature'', {\it The Astrophysical Journal} {\bf 835}, 26 (2017). 

\bibitem{refld0} J Rom\'an-Garza, T Verdugo, J Maga\~na and V Motta, ``Constraints on barotropic dark energy models by a new phenomenological $q(z)$ parameterization'', {\it The European Physical Journal C} {\bf 79}, 890 (2019).

\bibitem{Dinda:2019mev} BR Dinda, ``Model independent parametrization of the late time cosmic acceleration: Constraints on the parameters from recent observations'', {\it Physical Review D} {\bf 100}, 043528 (2019).

\bibitem{lin2019non} H-N Lin, X Li and L Tang, ``Non-parametric reconstruction of dark energy and cosmic expansion from the Pantheon compilation of type Ia supernovae'', {\it Chinese Physics C} {\bf 43}, 075101 (2019).

\bibitem{goswami2019friedmann} G Goswami, A Pradhan and A Beesham, ``Friedmann–Robertson–Walker accelerating Universe with interactive dark energy'', {\it Pramana} {\bf 93} 89 (2019).

\bibitem{jesus2019gaussian} J Jesus, R Valentim, A Escobal and S Pereira, ``Gaussian process estimation of transition redshift'', {\it arXiv}:1909.00090 (2019).

\bibitem{carvalho2008cosmological} F Carvalho, E Santos, J Alcaniz and J Santos, ``Cosmological constraints from the Hubble parameter on f(R) cosmologies'', {\it Journal of Cosmology and Astroparticle Physics} {\bf 2008}, 008 (2008).

\bibitem{dadhich2011enigmatic} N Dadhich, ``On the enigmatic $\Lambda$ -- a true constant of spacetime'', {\it Pramana} {\bf 77} 433 (2011).


\end{thebibliography}
\end{document}